\long\def\cut#1{}
\font\df=cmssbx10 
\def\cmp#1{}   

\documentstyle[preprint,aps]{revtex}

\def\mpm{{\bf m}_\pm}
\def\mpl{{\bf m}_+}
\def\mm{{\bf m}_-}
\def\m{{\bf m}}
\font\cmbsy=cmbsy10\def\bnabla{{\textfont2=\cmbsy\nabla}}  
\def\bold#1{\setbox0=\hbox{$#1$}%
     \kern-.010em\copy0\kern-\wd0
     \kern.025em\copy0\kern-\wd0
     \kern-.020em\raise.0200em\box0 }
\def\bep{\bold\varepsilon}
\def\KK{\hbox{\df K}}

\begin{document}
\begin{center}

\Large

{\bf The Role of Bilayer Tilt Difference in Equilibrium Membrane Shapes}

\large \vspace{0.5cm}
{\sl Udo Seifert$^1$,  Julian Shillcock$^1$, and  Philip Nelson$^2$}
\vspace{0.5cm}
\normalsize

$^1$Max-Planck-Institut f\"ur Kolloid- und
Grenzfl\"achenforschung, Kantstrasse 55, 14513 Teltow-Seehof, Germany;
%
%
$^2$Physics and Astronomy,
University of Pennsylvania, Philadelphia, PA 19104 USA
\end{center}
\begin {abstract}
Lipid bilayer membranes below their main transition have
two tilt order parameters, corresponding to the two monolayers. These
two tilts may be strongly coupled to membrane shape but
only weakly coupled to each other. We discuss some
implications of this observation for
rippled and saddle phases, bilayer tubules, and bicontinuous phases.
Tilt difference introduces a length scale into the
elastic theory of tilted fluid membranes. It can drive an instability
of the flat phase; it  also provides a simple  mechanism for the
spontaneous breaking of inversion symmetry seen in some recent
experiments.

\end{abstract}
\vskip 1cm
\def\beq{\begin{equation}}
\def\ee{\end{equation}}
\def\pcite{\protect\cite}
\def\pa{\partial}
\def\m{{\bf m}}
\def\q{{\bf q}}

\def\lsim {\protect
\raisebox{-0.75ex}[-1.5ex]{$\;\stackrel{<}{\sim}\;$}}

\def\gsim {\protect
\raisebox{-0.75ex}[-1.5ex]{$\;\stackrel{>}{\sim}\;$}}

\def\lsimeq {\protect
\raisebox{-0.75ex}[-1.5ex]{$\;\stackrel{<}{\simeq}\;$}}

\def\gsimeq {\protect
\raisebox{-0.75ex}[-1.5ex]{$\;\stackrel{>}{\simeq}\;$}}

The curvature model of fluid bilayer membranes has proved quite
successful in explaining the shapes of membranes above their main
transition~\cite{lipo91}. In this model locality, coordinate
invariance, and bilayer symmetry
restrict the form of the energy functional for shapes to just two
terms, involving the mean and Gauss
curvature. In fixed topology the total Gauss curvature is constant,
and so the minimum-energy conformation is a surface of vanishing mean
curvature, for example a flat plane.

Below the main transition, additional degrees of freedom enter the
elasticity of membranes as their hydrocarbon chains begin to order. In
analogy to smectic liquid crystals, one expects a soft {\it tilt}
degree of freedom to appear, reflecting the spontaneous breaking of
rotational invariance in the plane. Helfrich and Prost began the
systematic study of the mutual influence of tilt order and membrane
shape \cite{helf88}.
A number of nonflat ground state
phenomena found in membranes below their main transition have since been
attributed to tilt,  including rippled phases and tubule phases (see
for example  \cite{lube93,seli94,chen95,chen96}). Tilt order also
proves to be crucial for the intrinsic {\it chirality} of individual
amphiphiles (if any) to influence the 
conformations of membranes \cite{nels92}.

Despite much progress, however, a number of mysteries remain in the
study of one-component, symmetric bilayer membrane
conformations. For example,
experiments with racemic mixtures of lipids have found
asymmetric ripple ground states, which are chiral \cite{kats95}.
Similarly, achiral amphiphiles can form tubules \cite{sing88}, which
again appears to require chirality \cite{seli94}. Even chiral lipids easily
form helical ribbons of {\it either} handedness; sometimes a single
ribbon appears to switch
handedness in the middle of its growth \cite{thom94}. Finally, cubic
phases of bilayer
membrane are predicted to be scale-invariant in the pure curvature
model \cite{brui92}. One might therefore expect them to collapse to a
microscopic cell size, but in fact they can be stabilized at
well-defined, mesoscopic scales \cite{whoB}.
 Theoretically, selection of a
length-scale of several nanometers
has been attributed to higher order curvature terms both for
cubic phases \cite{brui92} and for a presumed hats and saddle super-structure
of fluid bilayers \cite{goet95}.

In this letter we  explore a new model for the 
conformation of membranes: we augment the curvature model with {\it
two} tilt director fields corresponding to  the two
monolayers. Thus our work fits into the general program of taking more
seriously the bilayer aspect of membranes, both in their equilibria
and dynamics (e.g.~\cite{yeun92,seif93,miao94}).
When the directors are aligned we reproduce existing
models; when they are not we get new physics.
Imagining that the
tilt in a monolayer induces spontaneous curvature orthogonal to its
direction, in the corresponding bilayer
the flat state with anti-parallel tilt then is  frustrated and thus
more costly than a saddle conformation in which the tilts are
oriented orthogonal to each other. Without interaction between the two
layers the flat state is always unstable locally against this kind
of saddle conformation.  Real membranes will always have at least some
tendency to align the tilts.
The instability will then occur only if
the anisotropic spontaneous curvature is larger than a threshold
value  determined below.

There seems to be no direct experimental evidence for or
against non-aligned tilt in bilayer phases.  We will motivate the
model, show how it gives a new mechanism for the transition from flat
to rippled membranes, and comment  on how it addresses the
questions above.

Our work was first motivated by a desire to understand the origin of
a nonanalytic curvature energy proposed by Fischer \cite{fish92}.
Our model is mathematically similar to one independently proposed
by Fournier \cite{four96}, but the physical motivation is quite
different: while he considered an anisotropic {\it impurity} adsorbed
onto a membrane, our tilt is an {\it intrinsic} property of a {\it
pure} bilayer, and hence quite generic. Other differences will be
noted below.

\medbreak{\noindent\sl Model:\ \ }We will restrict attention to
systems of nonchiral amphiphiles. In the covariant 
notation developed in \cite{nels92,nels93,powe95}, this means we
consider only elastic energy terms constructed without the
in-plane antisymmetric tensor $\epsilon_{\alpha\beta}$. To focus attention on
the new elements we will also impose an additional ``nematic''
symmetry (see below). This assumption is strictly for mathematical
simplicity; we leave the full model to future work.

Above the main transition we imagine the membrane to be two identical
2-dimensional fluid sheets of elastic monomers, independent except for
the constraint that they lie a fixed distance above and below a common
surface. We will label the layers arbitrarily as ``+'' and ``$-$'',
but since the layers are identical we will insist that nothing changes
if we reverse the labeling. To define the curvature tensor
$K_{\alpha\beta}$ 
we will choose the normal vector $\bf N$  pointing from the ``$-$'' to
the ``$+$'' side.
Each sheet has its own bending stiffness, stretching modulus,
and spontaneous curvature. When we combine the sheets, the bending
stiffnesses and stretching moduli add, while the spontaneous
curvatures cancel (for details see e.g.~\cite{miao94}). We will 
neglect the stretchiness of the membranes, leaving only the curvature
stiffness.

Below the main transition each monolayer develops a local average
tilt. We will not be interested in the main transition itself, so we
will take the director of the amphiphiles to be at a fixed 
angle to the layer normal, and the degree of ordering to be
constant. In other words, we will describe the tilt by a pair of
unit vector fields $\mpm$ tangent to the surface: these are the
normalized projections of the average molecular directors. The elastic
energy is then a local functional of $\mpl,\ \mm$, and the membrane
shape, described by its curvature tensor $K_{\alpha\beta}$. As
mentioned above, we will rather artificially assume invariance when
either $\mpl$ or $\mm$ changes sign.

Traditionally one takes $\mpl=-\mm$ (e.g. see \cite{nels93}), or in
other words assumes that the average directors in each layer are
collinear. The reasoning is that while an overall rotation of both
$\mpm$ is related to a broken symmetry, still the {\it relative} angle
is not, and so is expected to lock to a
preferred value. We propose to
explore what happens when this assumption is relaxed. Our motivation
is the observation that the degree of interdigitation of the lipid
chains between the layers is in fact quite small, as deduced for
example from measurements of the interlayer friction coefficient
\cite{yeun92}. We will allow for some weak aligning potential
$g((\mpl\cdot\mm)^2)$,
but as  we will show other energetic contributions
can readily overcome it. For simplicity we will continue to assume
that $g$ is minimized at $\mpl=-\mm$, but later on we will reconsider
this assumption too. In  ref.~\cite{four96}, $g$ was
taken to be zero, an extreme limit of our model.

While the direct interaction of the tilts may be small, there will
certainly be an important {\it indirect} interaction mediated by the
coupling of tilt to the common {\it shape}.  To lowest order in the
curvature tensor the effect of tilt on a monolayer is
to create an anisotropic spontaneous curvature, by adding the terms
\beq f^\pm=\beta\, K_{\alpha\beta}m_\pm^\alpha m_\pm^\beta\equiv
\beta\,\mpm\cdot\KK\cdot\mpm
\label{eq:ftwo}\ee
to the elastic free energy. $\beta$ is a new parameter depending on
the degree of ordering \cite{foot1}.
The combination $f^+-f^-$ is then invariant under renaming the two
leaves of the bilayer. It
vanishes
if $\mpl=-\mm$, but more generally
we need to keep these terms. At higher order in curvature we have the
more familiar terms $\kappa_3 (K_{\alpha\beta}m_\pm^{\alpha}m_\pm^{\beta})^2 +
\kappa_4 K_{\alpha\beta}K_{\beta\gamma}m_\pm^{\alpha}m_\pm^{\gamma}$ (see
\cite{powe95}). Since (\ref{eq:ftwo}) will already drive our
instability we will neglect the higher-curvature $\kappa_3,\ \kappa_4$
terms. All told, then, our simplified model is defined by the elastic
energy density functional
\begin{eqnarray}
&f=
{\kappa_1\over2} (K^{\ \alpha}_\alpha)^2 +
{\kappa_2\over2} K_{\alpha\beta}K^{\alpha\beta}
+ g((\mpl\cdot\mm)^2)\nonumber \\
&
+ {\displaystyle\sum_{\m=\mpm}}\left[ \pm  \beta
K_{\alpha\beta}m^{\alpha}m^{\beta}
+ {k_1\over4} \nabla_\alpha m^\beta
\nabla^\alpha m_\beta + {k_2\over4} (\nabla_\alpha m^\alpha)^2 \right].
 \label{eq:f}
\end{eqnarray}
The constants $\kappa_1,\ \kappa_2$ are related to the usual mean and
Gaussian rigidities,   while $k_1,\ k_2$ are related to the usual
rigidities of the XY model.

A remarkable feature of (\ref{eq:f}) is that  the parameter $\beta$
has dimensions of inverse length. This is a key qualitative
difference from the case of parallel tilt directors: achiral
symmetric bilayers with one tilt admit only {\it
dimensionless} couplings \cite{foot2,nels93}.
One
might expect that the value of such a parameter would be very large,
on the order of the microscopic wavenumber, and would create
correspondingly small structures. From its geometric origin, however,
it is easy to see how $\beta$ could be smaller than na\"\i ve
dimensional analysis would suggest. For example, the average tilt
could be rather small, leading to a small net spontaneous
curvature. In this way our term could set a mesoscopic scale.

\medbreak{\noindent\sl Stability of flat surfaces:\ \ }
To understand the physics of the model (\ref{eq:f}), we first note
that if the director in the ``+'' layer
is pointing in the $x$ direction, and the one in the ``$-$'' layer in the
$y$
direction, the $\beta$ term becomes
$\beta [K_{xx}-K_{yy}]$,
which favors  saddles (or other non-spherical shapes).

For a quantitative stability analysis of the flat state in which
both nematic fields are parallel, we need an explicit form of the
interaction $g$. A simple form  that favors parallel alignment is
$
g={\gamma} (1-(\mpl\cdot\mm)^2)/2 $.
We parametrize small deviations from this parallel state for the
director fields as $
{\m}_\pm=\left({\cos \phi}\atop{\pm\sin \phi}\right)\approx
\left({1- \phi^2/2}\atop{\pm  \phi}\right)  $.
In this representation, the coupling term between the two layers
becomes
$
g=2 \gamma \phi^2 $.

Adding up all energies to quadratic order in a Fourier representation for
$\phi(x,y)$
and
 the height $h(x,y)$  leads to
\begin{eqnarray}
f=&{\kappa\over 2} (\q^2)^2 h_\q^2+ 4 \beta q_xq_yh_\q\phi_\q+
2\gamma \phi_\q^2 \nonumber\\+& \left({k_1\over 2}(q_x^2+q_y^2)+{k_2\over
2}q_y^2\right)
\phi_\q^2
\end{eqnarray}
with $\kappa\equiv \kappa_1 + \kappa_2$.
Minimizing with respect to $h_\q$ yields
$
h_\q=-4\beta{q_xq_y\over\kappa(\q^2)^2}\phi_\q $.
Inserting this result into (\ref{eq:f}) leads to  the effective energy for
the $\phi$ field as
\beq
f= \left[{{ -8 \beta^2q_x^2q_y^2\over\kappa(\q^2)^2}} + 2\gamma+
{k_1\over 2}(q_x^2+q_y^2)+{k_2\over 2}q_y^2\right] \phi_\q^2 \ .
\ee

The stability criterion for the flat phase is now obvious. For weak
enough interlayer coupling,
\beq
\gamma<\beta^2/\kappa  ,
\ee
the flat phase becomes unstable to a long-wavelength modulation.
This is our main result.
The preferred directions for the $q$ vectors are
$q_x=\pm q_y=\pm\sqrt{\q^2}/2$.
Adding just two modes with $\q_1=-\q_2$ yields a ripple shape
whereas adding  four modes with $\pm q_x=\pm q_y$ leads to a egg-carton-like
square modulated phase.

Lubensky and MacKintosh also obtained
symmetric ripples in a nonchiral model. Then balanced an effectively
negative $(\nabla \m)^2$ term
against a stabilizing $(\nabla^2\m)^2$ term \cite{lube93}. Far from
the main transition
such a balance is likely to select a microscopic length scale. In
contrast, we have seen how our model can select long lengths. Other
models assumed hexatic order \cite{lube93,chen95}, while we have not.


\def\bea   {\begin{eqnarray}}
\def\eea   {\end{eqnarray}}

\def\mh  {h(x,y)}
\def \phip {\phi^{+}}
\def\phim {\phi^{-}}
\def\phipx {\phi^{+}(x,y)}
\def\phimx {\phi^{-}(x,y)}
\def\bt {{\beta}}

\medbreak{\noindent\sl Beyond instability:\ \ }
In order to distinguish the two alternatives, ripple or saddle,
we have to go beyond the instability. We will do so in the
following using exact analysis, a simple  variational shape
and numerical minimization.
To keep our formulas tractable, in this section we will neglect the
tilt stiffness terms, i.e. we set $k_1= k_2=0$. Thus the tilt fields
track the curvature exactly, and in particular can change discontinuously
by $\pi/2$ when the mean curvature changes sign. A more realistic
model would broaden these discontinuities into domain walls, with an
energy cost per unit length.

First, we discuss the ripple phase. For a one-dimensionally modulated
conformation, the free energy density takes the form
\begin{eqnarray}
f = & (\kappa/2)(h_{xx})^2 + \beta h_{xx}
 (\cos^2\phip - \cos^2\phim)\nonumber\\
&+{\gamma\over 2} (1-(\cos\phip\cos\phim +
\sin\phip\sin\phim)^2) .\label{rippleh}
\end{eqnarray}
One could try as a trial variational function a sinusoidal ripple with
wavenumber $q$ and height $h_q$ and nematic fields always arranged
perpendicular to each other which leads to an energy density
of $f/\kappa = -{1\over 2} (\beta^2 {8 \over \pi^2} -{\gamma})$. However
one can easily see  that the sinusoidal shape
is far from optimal. If the tilts are orthogonal, thus rendering
the  $\beta$ term $- |\beta h_{xx}| $, ripples that consist of
circular arcs of radius $\kappa/\beta$ are lower in energy than
sinusoidal ripples.
A straightforward calculation shows  that allowing the fields to
assume a non-orthogonal configuration does not lead to
a lower energy for  $\gamma<\beta^2/\kappa$ which is the region
of interest. Thus, in summary, as long as only one-dimensionally
modulated phases are considered, a ripple phase
consisting of circular arcs with the nematic
fields orthogonal in the two layers  is the most
favorable configuration. Due to  the absence of gradient terms
 for the director fields, these fields jump whenever the
curvature changes sign. Including the gradient terms would thicken the
jump region to a domain wall; the energy of this wall would then
favor long wavelength (fewer walls per unit length), breaking the
degeneracy we have found so far.

When modulation of the membrane in two dimensions is allowed, the free
energy density  takes the form
\bea
&f =(\kappa/2) (h_{xx} + h_{yy})^2  +(\gamma/2) (1 -
\cos^2(\phip - \phim)) \nonumber \\
     &+  \beta\bigl[h_{xx}(\cos^2\phip -\cos^2\phim)
              + h_{yy}(\sin^2\phip - \sin^2\phim) \nonumber\\
&\qquad               + h_{xy}(\sin2\phip - \sin2\phim) \bigr].
\label{eq:f2}         \eea
Again a variational trial function sinusoidal in two directions is not
the best choice. It leads to an energy density
of $f/\kappa= -{1\over 2} (\beta^2 {64\over \pi^4} -\gamma)$.
Using the fact that the lowest
energy ripple conformation had constant curvature, we
construct a saddle phase in which the curvature is piecewise constant.
This leads to a saddle conformation in which the
corrugations are parabolic arcs.  In
regions where the principal curvatures have the same sign, we
allow the nematic
fields to be parallel (thus gaining from the $\gamma$ term),
and where they have opposite signs the fields
remain orthogonal. The free energy density of this state is
$
f/\kappa = -{1\over 4} (\beta^2 - \gamma)$. 
This energy is never lower than either the flat phase or the
circular arc ripples even though it improves the value of the
sinusoidal saddle for a large range of $\gamma$ values.

Having exhausted the range of analytic saddle conformations, we turned
to a numerical minimization of the full free energy (\ref{eq:f2}).
While we could improve on the parabolic
saddles,  still we found no saddle phase
with lower energy than either the circular ripples or the flat phase.

\medbreak{\noindent\sl Discussion:\ \ }
We have not obtained a unique
wave-length for the instability. What we found was rather that the
tilt-difference coefficient $\beta$ sets a preferred {\it
curvature} for cylindrical segments. For small amplitudes this means
 that the combination $h_qq^2$,
but not $q$ itself, is fixed. This happened because
we neglected the gradient terms in the tilt fields:
the  energy of a configuration is then
invariant under a rescaling of $h(x,y) \to \lambda h (\lambda x,
\lambda y)$ if at the same time $\gamma\to \gamma/\lambda^2$ and $\beta
\to \beta/\lambda$. Gradient terms in $\m$, of course, would favor
large wave-length, up to a maximum of $\sim\kappa/\beta$, where the
membrane rolls up into cylinders.
One way to stabilize finite-wavelength ripples would be to stretch the
membrane with a lateral tension.

Alternatively we can take our results as indicating an instability to
a phase of cylinders or other curved objects, especially saddles.
Particularly intriguing
is the possibility that the tilt-difference term could set a scale for
bicontinuous phases without resorting to microscopic
(higher-curvature) terms as in \cite{brui92}. \cut{As mentioned in the
Introduction, some physics
sets a mesoscopic cell size in recent experiments. Our $\beta$ term
can do this because, after integrating over a unit cell, its value
scales differently with cell size from the Gauss curvature term.}

We now consider briefly the possibility of breaking
spatial inversion symmetry.
Other authors have obtained rippled or striped phases in tilted
bilayers by explicitly breaking parity invariance
(e.g. \cite{lube93,seli94}). In  particular, explicit
parity-violating terms in the free energy have so far been necessary
to obtain {\it asymmetric}, or sawtooth, rippled phases. But the
presence of chiral amphiphiles does not guarantee a parity-breaking
Landau energy \cite{kami96}. Moreover as mentioned earlier, several \cut{in the
Introduction, chiral structures form even from achiral amphiphiles;
moreover even chiral amphiphiles do not always determine the handedness of the
chiral structures they form. All these} arguments point to the possibility
of {\it spontaneous breaking} of parity invariance: the Landau energy
has parity invariance, but its {\it minima} do not.

In an elegant paper, Selinger {\it et al.}\ have proposed possible
mechanisms for
spontaneous parity breaking in monolayers \cite{seli93}. Two of these
could also apply to  pure bilayers: (a)~if the membrane has
both tilt and hexatic order, the corresponding directors may lock to a
fixed, nonzero {\it relative} angle; (b)~conceivably two distinct local
packings of molecules could be preferred, each of which is the other's
mirror image. We would like to point out a very simple, concrete
option with some elements of each of these: (c)~a tilted membrane may
prefer on packing grounds to order its two tilt directors at a fixed
relative angle $\alpha=\pm\alpha_0$, i.e.\
$\mpl\cdot\mm=\cos\alpha_0$.  This option
may prove more generic than 
the ones above.

To see how our proposal leads to parity breaking,  note that
$\psi\equiv (\mpl\times\mm)\cdot{\bf N}$ is a well-defined pseudoscalar order
parameter. $\psi$ vanishes as
$\alpha\to0$, as required.

We can write
$\mpm=\left[\begin{array}{cc}
\cos\alpha/2&\pm\sin\alpha/2\\\mp\sin\alpha/2&\cos\alpha/2\end{array}\right]
\m$, where
$\m$ is a  common tilt variable.
Writing {\df1} for the unit matrix and
$\bep=\left[ \begin{array}{cc}0&1\\-1&0\end{array}\right]$,
the terms (\ref{eq:ftwo}) now read
\bea
f^+-&f^-=\beta\m\cdot\bigl[
(\cos{\alpha\over2}\hbox{\df1}-\sin{\alpha\over2}\bep)
\KK(\cos{\alpha\over2}\hbox{\df1}+\sin{\alpha\over2}\bep)
\nonumber\\&-
(\cos{\alpha\over2}\hbox{\df1}+\sin{\alpha\over2}\bep)
\KK(\cos{\alpha\over2}\hbox{\df1}-\sin{\alpha\over2}\bep)
\bigr]\cdot\m\ .
\eea
Rewriting this as $(2\beta\sin{\alpha\over2})\m\cdot[\bep
\KK]\cdot\m$, we recognize the
chiral tilt-shape coupling introduced by Helfrich and Prost
\cite{helf88}, in the coordinate-invariant form given in
\cite{nels93}. This is precisely the term
responsible for the formation of asymmetric ripples in the work of
Lubensky and MacKintosh \cite{lube93}. A similar analysis, dropping
the requirement of nematic symmetry, could also yield the chiral term
which gives rise to tubules and helices \cite{seli94}.

In conclusion, we have shown that tilt difference may be expected
quite generally to
affect the conformations of symmetric bilayer membranes, introducing a new
intermediate length scale and favoring cylindrical and saddle curvature over
flat or spherical shapes. Tilt difference could also provide an attractively
general mechanism for the spontaneous breaking of parity invariance.

\medbreak{\noindent\sl Acknowledgements:\ \ }
We are grateful to T. Fischer, R. Kamien, T. Lubensky, and J. Prost
for helpful discussions.
This work was supported in part by the
US/Israeli Binational Foundation grant 94--00190 and NSF grant
DMR95--07366.


\begin{thebibliography}{10}

\bibitem{lipo91}
R. Lipowsky, \cmp{``The conformation of membranes,''} Nature
{\bf349} (1991) 475.

\bibitem{helf88}
W. Helfrich and J. Prost, \cmp{``Intrinsic bending force in anisotropic
membranes made of chiral molecules,''}
Phys. Rev. {\bf A38} (1988) 3065.

\bibitem{lube93}
T. Lubensky and F. MacKintosh, \cmp{``Theory of `ripple' phases of lipid
bilayers,''} Phys. Rev. Lett. {\bf71} (1993) 1565. 

\bibitem{seli94}
J. Selinger and J. Schnur, \cmp{``Theory of chiral lipid tubules,''}
Phys. Rev. Lett. {\bf 71} (1993) 4091; 
J. V. Selinger, F. C. MacKintosh, and J. M. Schnur, \cmp{``Theory of
Cylindrical Tubules and Helical Ribbons of
   Chiral Lipid Membranes,''} Phys. Rev. {\bf E53} (1996) 3804.

\bibitem{chen95}
C.-M. Chen, T. Lubensky, and F. MacKintosh, \cmp{``Phase transitions and
modulated phases in lipid bilayers,''} Phys. Rev. {\bf E51} (1995)
504. 

\bibitem{chen96}
C.-M. Chen and F. MacKintosh, \cmp{``Theory of modulated phases in lipid
bilayers and liquid crystal films,''}
Phys. Rev. {\bf E53} (1996) 4933. 

\bibitem{nels92}
P. Nelson and T. Powers, \cmp{``Rigid Chiral Membranes,''}
Phys. Rev. Lett. {\bf69} (1992) 3409.


\bibitem{kats95}
J. Katsaras and  V. Raghunathan,
    \cmp {``Molecular chirality and the ripple phase of phosphatidylcholine
   multibilayers,''}
     Phys. Rev. Lett.  {\bf 74} (1995) 2022. 

\bibitem{sing88}
A. Singh, P. Schoen, J. Schnur, Chem. Commun. {\bf
1988} (1988) 1222.

\bibitem{thom94}
B. Thomas, talk presented at \cmp{``Biomolecular Materials,''} Santa Barbara
(Sept. 1994); B. Thomas, in preparation.

\bibitem{brui92}
R. Bruinsma,  \cmp{``Elasticity and excitations of minimal crystals,''}
    J. Phys. II (France) {\bf 2} (1992) 425. 

\bibitem{whoB}
U. Peter, S. K\"onig, D. Roux, and A.-M. Bellocq,
 Phys. Rev. Lett.  {\bf 76} (1996) 3866.

\bibitem{goet95}
R. Goetz and W. Helfrich,  J. Phys. II (France)  {\bf 6} (1996) 215, and
references therein.

\bibitem{yeun92}
E. Evans, A, Yeung, R. Waugh, and J. Song, {\cmp{``Dynamic
coupling and nonlocal curvature elasticity in bilayer membranes,''}} in
{\sl Structure and conformation of amphiphilic membranes,} ed. R.
Lipowsky  (Springer, 1992); E. Evans and A. Yeung,
\cmp{``Hidden dynamics in rapid changes of bilayer
shape,''} Chem. Phys. Lipids {\bf73} (1994) 39. 


\bibitem{seif93}
U. Seifert and S. Langer, \cmp{``Viscous modes of bilayer
membranes,''} Europh. Lett. {\bf23} (1993) 71; 
\cmp{``Hydrodynamics of membranes: the bilayer aspect and adhesion,''}
Biophys. Chem. {\bf49} (1994) 13. 

\bibitem{miao94}
L.
Miao, U. Seifert, M. Wortis, and H.-G. D\"obereiner, {\cmp{``Budding
transitions of fluid-bilayer vesicles,''}} Phys. Rev. {\bf E49} (1994)
5389. 

\bibitem{fish92}
T. Fischer, J. Phys. II (France) {\bf2} (1992) 337; {\it ibid.} {\bf3}
(1993) 1795.

\bibitem{four96}
J. Fournier, \cmp{``Nontopological saddle-splay and curvature instabilities
from anisotropic      membrane inclusions,''}
   Phys. Rev. Lett. {\bf76} (1996) 4436. 

\bibitem{nels93}
P. Nelson and T. Powers, \cmp{``Renormalization of chiral
couplings in tilted bilayer membranes,''} J.  Phys.  II (France)  {\bf3}
(1993) 1535. 

\bibitem{foot1}
The term $(K_\alpha^{\ \alpha})
(\bnabla\cdot\m)$, which drove the rippling transition in
\cite{lube93}, does not have our nematic $\m\to-\m$ symmetry. We
will obtain an instability even without this term. Our term
(\ref{eq:ftwo}) is reminiscent of a coupling to {\it composition} difference
\cite{lei87}, but with a tensor, not scalar, character. Again we
emphasize that our model deals with pure, one-component membranes.

\bibitem{lei87}
S. Leibler and D. Andelman, J. Phys. (Paris) {\bf48} (1987) 2013.


\bibitem{powe95}
T. Powers and P. Nelson, \cmp{``Fluctuating Membranes With Tilt Order,''}
J. Phys. II
(France) {\bf5} (1995) 1671. 

\bibitem{foot2}
This result is valid even if we do
not impose the artificial nematic symmetry used here.

\bibitem{kami96}
A.B. Harris, R.D. Kamien
and T.C. Lubensky, \cmp{``On the Microscopic Origin of Cholesteric Pitch,''}
preprint 1996.

\bibitem{seli93}
J. Selinger, in {\sl Complex Fluids,}\/ ed. E. Sirota, D. Weitz,
T. Witten, and J. Israelachvili (MRS, 1992), p.~29;
J. Selinger, Z.-G. Wang, R. Bruinsma, and C. Knobler, \cmp{``Chiral
symmetry breaking in Langmuir monolayers and smectic films,''}
Phys. Rev. Lett. {\bf70} (1993) 1139. 


\end{thebibliography}
\end{document}